\documentclass[11pt]{article}
\usepackage{amsmath,amsthm,amssymb}
\usepackage{dsfont}
\usepackage{graphicx}
\usepackage{epsfig}
\usepackage{natbib}
\usepackage{algorithm}
\usepackage{algpseudocode}
\usepackage{hyperref}
\usepackage{xcolor}
\usepackage{url}            
\hypersetup{
  colorlinks = false
}

\usepackage[margin=1.25in]{geometry}

\newtheorem{theorem}{Theorem}
\newtheorem{condition}{Condition}
\newtheorem{lemma}{Lemma}

\def\EE{\mathbb E}

\def\RR{\mathbb R}

\def\bth{\boldsymbol{\theta}}

\begin{document}
\title{A Nonparametric Bayesian Approach for Sparse Sequence Estimation}

\author{Ouyang Yunbo, Feng Liang \\
\small Department of Statistics \\ [-0.8ex]
\small University of Illinois at Urbana-Champaign \\ [-0.8ex]
\small Champaign, IL, USA \\ [-0.8ex]
\small \texttt{youyang4@illinois.edu}
}

\maketitle

\begin{abstract}
A novel Bayes approach is proposed for the problem of estimating a sparse sequence based on Gaussian random variables. We adopt the popular two-group prior with one component being a point mass at zero, and the other component being a mixture of Gaussian distributions. Although the Gaussian family has been shown to be suboptimal for this problem, we find that Gaussian mixtures, with a proper choice on the means and mixing weights, have the desired asymptotic behavior, e.g., the corresponding posterior  concentrates on balls with the desired minimax rate. To achieve computation efficiency, we propose to obtain the posterior distribution using a deterministic variational algorithm. Empirical studies on several benchmark data sets demonstrate the superior performance of the proposed algorithm compared to other alternatives.
\end{abstract}


\section{Introduction}
Consider a Gaussian sequence model
\begin{equation} \label{eq:model}
X_i = \theta_i + e_i, \quad i=1, 2, \dots, n,
\end{equation}
where $\bth= (\theta_1, \dots, \theta_n)$ is the unknown mean parameter, often assumed to be sparse, and $e_i$'s are independent errors following a standard normal distribution. The primary interest is to reconstruct the sparse vector $\bth$ based on the data $X^n = (X_1, X_2, \cdots, X_n)$. Such a simple model arises in many applications, such as astronomy, signal processing and bioinformatics. It is also a canonical model for many modern statistical methods, such as nonparametric function estimation, large-scale variable selection and hypothesis testing.

There are different ways to define the sparsity of a vector. In this paper, we focus on the common definition of sparsity, i.e., many elements of $\theta_i$'s are zero. In particular, we assume $\bth$ is from the class of \emph{nearly black} vectors,
\[ \Theta_0(s_n) = \Big \{ \bth \in \RR^n: \# \{1 \le i \le n: \theta_i \ne 0 \} = s_n \Big \}. \]
The parameter $s_n$ measures the sparsity of $\bth$, which is usually assumed to be $o(n)$, but unknown. So a desired feature of any reconstruction procedure is to adapt to the unknown sparsity level.

Since $\bth$ is known to be sparse, a natural approach is to threshold $X^n$. The thresholding rule can be chosen by the principle of minimizing the empirical fitting error with penalization \citep{golubev:2002} or by minimizing the False Discover Rate \citep{abramovich2006special}. In addition, a plethora of research on variable selection and prediction in the context of high dimensional linear regression models can  also be applied on this problem\citep{fan2010selective, buhlmann2011statistics, zhang2008sparsity}.

Some thresholding procedures are motivated from a Bayesian aspect, such as \cite{george2000calibration},  \cite{johnstone2004needles}, and \cite{castillo2012needles}. The idea is to model $\theta_i$'s with a two-group structure prior,  where one component is a point mass at zero due to the sparsity assumption and the other component is a continuous distribution with a smooth symmetric density function $g$, namely,
\begin{equation} \label{prior:two-group}
\pi(\theta_i) = w \delta_0(\theta_i) + (1- w) g(\theta_i).
\end{equation}
Theoretical studies
have indicated that $g(\cdot)$, the prior density function on the non-zero component, should have a heavy tail for the posterior distribution to have the desired asymptotic behavior. Therefore, the double exponential distribution and its scaled variants are recommended for the choice of $g$, but not Gaussian distributions. The problem with distributions like the Gaussians which have lighter tails is that they tend to over-shrink $X_i$'s for large $\theta_i$'s therefore attain a lower posterior contraction rate  \citep{johnstone2004needles, castillo2012needles}.

In this paper, we propose an Empirical Bayes approach to this problem. Despite the warning message on Gaussian distributions, we still use Gaussians in our prior specification, which  takes the following form
\begin{equation} \label{prior}
\pi(\theta_i) = w \delta_0(\theta_i) + (1- w) \sum^T_{t=1}w_t\textsf{N}(m_t, \sigma^2),
\end{equation}
then use data to learn weights $w, w_1, \cdots, w_T$ and centers $m_1, \cdots, m_T$. Our prior choice can be viewed as a special case of the two-group prior (\ref{prior:two-group}) with $g$ being a mixture of Gaussian density functions. The Gaussian prior is appealing due to its conjugacy, which simplifies our analysis and also enables tractable computation. As revealed by our asymptotic analysis, the suboptimal behavior of Gaussian distributions as mentioned in \cite{johnstone2004needles} and  \cite{castillo2012needles} is on Gaussian distributions with mean zero, which can be avoided by a proper choice on weights $w_1, \cdots, w_T$ and centers $m_1, \cdots, m_T$.

The remainder of this paper is organized as follows. We present our main results in Section 2. We show that with some mild conditions on $m_t$'s and $w_t$'s, the corresponding posterior will have desired asymptotic behavior: it concentrates around the true parameter $\bth^*$ at the minimax rate, its effective dimension adapts to the unknown sparsity $s_n$, and the corresponding the posterior mean is asymptotic minimax.  A variational implementation of our approach, as well as empirical studies, is presented in Section 3.
All the proofs are given at the end after the conclusion.\\

We will start the remaining of this paper  by briefly discussing some related work.
\begin{itemize}
\item This paper is motivated by a recent work by \cite{martin:walker:2014}, where $g(\theta) = g_i(\theta)$ in \eqref{prior:two-group} is set to be a Gaussian distribution centered exactly at the data point $X_i$. In  \cite{martin:walker:2014}, the resulting estimate of $\theta_i$ is still a shrinkage estimate toward zero, while in our approach, the estimate of $\theta_i$ is
adaptively shrunk toward the mean of nearby data points. This explains why the empirical performance of our approach is better than the one from \cite{martin:walker:2014}.

\item Estimating $\bth$ for the Gaussian sequence model \eqref{eq:model} can be also viewed as a compound decision problem \citep{robbins:1951}, where  $\theta_i$'s are assumed to be i.i.d. random variables from a common but unknown distribution $G$. The aforementioned two-group prior (\ref{prior:two-group}) can be viewed as a special form of $G$, but in general $G$ can take any form. A number of nonparametric approaches have been proposed to estimate $G$ and then in return provide an estimate for $\bth$, such as the Maximum likelihood approach by  \cite{Jiang:Zhang:2009}, the nonparametric empirical Bayes approach by \cite{Brown:Green:2009}, and a convex optimization based approach by \cite{koenker:bakeoff:2014}. In particular, Gaussian mixtures are used to estimate $G$ by \cite{Jiang:Zhang:2009} and \cite{Brown:Green:2009}. However, their asymptotic results do not cover the nearly black class $\Theta_0(s_n)$. In their simulation study, \cite{Jiang:Zhang:2009} indeed consider the sparse situation and suggest to add a Gaussian component with mean zero. In our asymptotic study, we have found that to achieve the desired asymptotic properties, a point mass zero, instead of a Gaussian with mean zero, seems necessary for the nearly black class.
\end{itemize}

\section{Main Results}
\label{mainresult}
First we introduce some key notations. Data vector is denoted as $X^n=(X_1,X_2,\cdots,X_n)$; $\Pi_{X^n}(\cdot)$ is used to denote the prior indicating dependence with data; $\bth^\ast=(\theta^\ast_1,\cdots,\theta^\ast_n)$ is the unknown true mean of $X^n$; $p_{\bth}(X^n)$ is the likelihood function of $\bth$ given $X^n$. $a_n \succeq b_n$ is used to represent $\lim_n b_n/a_n = 0$.

Consider the following hierarchical prior  $\Pi_{X^n} (\cdot \mid w_{1:T}, m_{1:T}, \sigma^2, \alpha)$ on $\bth$ as follows
\begin{align}\label{prior:new}
w &\sim \textsf{Beta}(\alpha n,1),\\
\theta_i \mid w & \sim w \delta_0 +(1-w)\sum^T_{t=1}w_t\textsf{N}(m_t,\sigma^2), i=1,\cdots,n
\end{align}
where $\sigma^2$, $\alpha$, $w_{1:T}$ and $m_{1:T}$ are fixed parameters. $T$ is independent of the dimension $n$.

The posterior distribution is proportional to $p_{\bth}(X^n)\Pi_{X^n}(\bth)$. Strong assumptions are usually needed to prove Bayesian consistency. However, following \cite{Walker:Hjort:2001} and \cite{martin:walker:2014}, we can obtain our posterior distribution with fractional likelihood, which can weaken the conditions required for Bayesian consistency. Given a fractional parameter $\kappa$, the corresponding posterior distribution is then proportional to $p_{\bth}^\kappa(X^n)\Pi_{X^n}(\bth)$. In implementation, $\kappa$ is set to be a large number close to 1 to capture most information from data. In all of the simulation studies in Section 3, we set $\kappa=0.99$. In particular, the posterior measure of a Borel set $A \subset \mathbb{R}^n$ involving $\kappa$,  denoted by $Q_n(A)$,  can be expressed as
\begin{equation}
 Q_n(A) = \frac{\int_A  \{p_{\bth}(X^n)/p_{\bth^\ast}(X^n)\}^\kappa \Pi_{X^n}(d\bth)}{\int_{\RR^n} \{p_{\bth}(X^n)/p_{\bth^\ast}(X^n)\}^\kappa \Pi_{X^n}(d\bth)}.
\end{equation}

$\bth^\ast$ is assumed to be sparse. Denote $S^\ast$ as the support of $\bth^\ast$ and $s_n= |S^\ast|$ is the cardinality. Minimax rate established in \citep{donoho:etla:1992} for estimating a vector in $\Theta_0(s_n)$ is $\varepsilon_n=s_n\log(n/s_n)$. The aim is to prove posterior mean estimator based on the above prior attains asymptotically minimax $L^2$ error rate in $\Theta_0(s_n)$. To accomplish this, first we prove our posterior measure $Q_n$ has a desired concentration rate and then prove MSE is bounded.

The key conditions are imposed on $w_1,\cdots, w_T$ and $m_1, \cdots, m_T$. To find suitable $m_1, \cdots, m_T$, we apply suitable clustering algorithms on nonzero mean set $\bth_{S^{\ast}}=\{\theta_i^{\ast}|i\in S^{\ast}\}$ and retrieve cluster centers as $m_1, \cdots, m_T$. Condition 1 essentially puts an upper bound on within-cluster sum of squares of $\bth_{S^{\ast}}$ if we apply some clustering algorithm. Denote $m_{t_i}$ as the corresponding cluster center of $\theta_i^{\ast}\in \bth_{S^{\ast}}$, we assume
\begin{condition}
$\sum_{i \in S^\ast} (\theta_i^\ast-m_{t_i})^2=o(\varepsilon_n)$.
\end{condition}
Since the summation in the left hand side has $s_n$ terms, if $\max_{i \in S^\ast} |\theta_i^\ast-m_{t_i}|=o(\log(n/s_n))$, that is, the distance between $\theta_i^\ast$ and the corresponding cluster centers grows slower than $\log(n/s_n)$, our condition could be satisfied. Therefore Condition 1 is not strict since $s_n=o(n)$. In implementation to specify $m_1, \cdots, m_T$, since we do not know $\bth_{S^{\ast}}$, we use a novel clustering algorithm to identify the cluster centered at 0 and estimate $T$ nonzero cluster centers. The details will be provided in Section \ref{implementation}.

The second condition is on the weight $w_1, w_2, \cdots, w_T$:
\begin{condition}
$\min_{1\leq t\leq T} w_t \geq C s_n/n>0$.
\end{condition}
Since $T$ is independent of $n$. This condition could be easily satisfied if we simply set $w_t=1/T$ for each $t$. If each cluster size is bounded below, we could also plug in cluster weights.

Given the above 2 conditions, first we introduce Lemma 1, an analogue of Lemma 1 in \cite{martin:walker:2014}.
\begin{lemma} \label{lemma:denominator}
Let $D_n$ be the denominator of posterior measure $Q_n$. If Condition 1 and Condition 2 hold, then $D_n>\frac{\alpha}{1+\alpha}\exp\{-2\varepsilon_n-o(\varepsilon_n)\}$ with $P_{\bth^\ast}-$probability 1.
\end{lemma}

After establishing a lower bound for the denominator of $Q_n$, in order to prove posterior concentration, we'll establish an upper bound for that numerator of $Q_n$ measuring the complement of a ball centered at $\bth^\ast$.
Specifically, we  show that posterior probability measure $Q_n$ concentrates asymptotically on a ball centered at the truth $\bth^\ast$ with square radius proportional to $\varepsilon_n$, namely,
$$ A_{M\varepsilon_n}=\{\bth\in \RR^n:\|\bth -\bth^\ast\|^2>M\varepsilon_n\}.$$

\begin{theorem} \label{thm:concentration}
If Condition 1 and Condition 2 hold, then there exists $M>0$ such that $Q_n(A_{M\varepsilon_n})\rightarrow 0$ with $P_{\bth^\ast}$-probability 1 as $n, s_n \to \infty$ with $s_n = o(n)$.
\end{theorem}

 Theorem \ref{thm:concentration} implies that  our posterior distribution concentrates around the right place at the right rate, so it ought to produce an estimator of $\bth$ with good properties. Next we show that the posterior mean $\hat{\bth}$ is a minimax estimator.

\begin{theorem} \label{thm:postmean}
If Condition 1 and Condition 2 hold, there exists a universal constant $M'>0$, such that $\EE_{\bth^\ast}\|\hat{\bth}-\bth^\ast\|^2\leq M'\varepsilon_n$ for all large $n$.
\end{theorem}

The proof is the same as the one by \cite{martin:walker:2014}, provided that we have proved Lemma \ref{lemma:denominator} and Theorem \ref{thm:concentration}.

As posterior concentrates around $\bth^\ast$ at the minimax rate, we could conclude the majority of the posterior mass concentrates on $s_n$-dimensional subspaces of $\RR^n$. The sparsity level of posterior is measured by the posterior distribution of $w$. Theorem \ref{th3} shows posterior distribution of $w$ put large probability mass above $1-s_nn^{-1}$.

\begin{theorem}\label{th3}
If Condition 1 and Condition 2 hold, let $\delta_n=K\varepsilon_nn^{-1}$, and $K>0$ is a suitably large constant. Then $E_{\bth^\ast}\{P(1-w>\delta_n|X^n)\}\rightarrow 0$ as $n\rightarrow \infty$.
\end{theorem}

Notice that $\delta_n=Ks_n/n\log(n/s_n)\approx Ks_n/n$, therefore the posterior distribution of $w$ concentrates around $1-s_nn^{-1}$. That is, the effective dimension of our posterior distribution adapts to the unknown sparsity level $s_n$. Our posterior mean estimator could detect true sparsity pattern with large probability. The proof is the same as the one by \cite{martin:walker:2014} as long as we have proved Lemma \ref{lemma:denominator} and Theorem \ref{thm:concentration}.

\section{Implementation}
\label{implementation}

In Section \ref{mainresult} we have shown using good prior with parameters $(w_{1:T}, m_{1:T}, \sigma^2, \alpha)$ could result in posterior concentration and minimax rate of Bayes posterior mean estimator. In implementation, we first specify $\sigma^2$ and $\alpha$, then
estimate $(w_{1:T}, m_{1:T})$ using $X^n$. In Condition 1 $\sum_{i \in S^\ast} (\theta_i^\ast-m_{t_i})^2$ needs to be bounded, however, in practice $\bth_{S^\ast}$ is unknown so that we cannot apply clustering algorithms on it. One remedy is to apply the clustering algorithm on $X^n$ since $\sum_{i \in S^\ast} (\theta_i^\ast-m_{t_i})^2\leq 2\sum_{i \in S^\ast} (X_i-m_{t_i})^2 + 2\sum_{i \in S^\ast} (X_i-\theta_i^\ast)^2 $. $\sum_{i \in S^\ast} (X_i-\theta_i^\ast)^2 = o_p(\varepsilon_n)$, therefore $\sum_{i \in S^\ast} (\theta_i^\ast-m_{t_i})^2 = o(\varepsilon_n)$ if $\sum_{i \in S^\ast} (X_i-\theta_i^\ast)^2 = o_p(\varepsilon_n)$.

 we apply some proper clustering algorithm on $X^n$ to assign clustering centers to $m_{1:T}$ and cluster weights to $w_{1:T}$.
 We need to pre-specify 0 as one clustering center so that the major task is to estimate the nonzero cluster cneters. In Bayesian Framework, Dirichlet process mixture model is used to do clustering. We build a Dirichlet process  (DP) mixture model on $X^n$ and estimate $m_t$ by plugging in corresponding cluster centers. The general framework is summarized as follows:
\begin{align*}
\theta_i&\sim G, G\sim \textsf{DP}(\alpha_0,G_0);\\
X_i&\sim \textsf{N}(\theta_i,1), i=1,2,\cdots, n;
\end{align*}
where $G_0$ is the base measure and $\alpha_0$ is the concentration measure. Given $G_0$ and $\alpha_0$,  we have stick breaking representation of $G$ as $\sum^\infty_{t=1}\pi_t\delta_{\eta_t}(\cdot)$, where $\eta_t$ is drawn independently and identically distributed (i.i.d.) from base measure $G_0$, while $\pi_t=V_t\prod^{t-1}_{l=1}(1-V_l)$. $V_l$ is drawn i.i.d. from $\text{Beta}(1,\alpha)$. From this formulation we could see that random distribution function $G$ is almost surely discrete. Since $\bth^\ast$ is a sparse vector, in order to get a random distribution $G$ with a positive point mass at 0, $G_0$ ought to have a positive mass at 0. Therefore we model $G_0$ as a normal component with a point mass at 0:
\[G_0=w_0\delta_0+(1-w_0)\textsf{N}(0,\sigma_0^2);\]
where $w_0$ and $\sigma_0^2$ are 2 pre-specified parameters. Then $G$ could be written as
\begin{align*}
&G=\sum^\infty_{t=1}\pi_t(\xi_t\delta_0(\cdot)+(1-\xi_t)\delta_{\eta_t^\ast}(\cdot))=\sum^\infty_{t=1}\pi_t\xi_t\delta_0(\cdot)+\sum^\infty_{t=1}\pi_t(1-\xi_t)\delta_{\eta_t^\ast}(\cdot)\\
&\equiv w\delta_0(\cdot)+(1-w)\sum^\infty_{t=1}w_t\delta_{\eta_t^\ast}(\cdot),
\end{align*}
where $w=\sum^\infty_{t=1}\pi_t\xi_t$ and $w_t=\frac{\pi_t(1-\xi_t)}{1-w}$. In this formulation
$\xi_t$ is drawn i.i.d. from $\textsf{Ber}(w)$. If $\xi_t=0$, then $\eta_t^\ast$ is drawn from $\textsf{N}(0,\sigma_0^2)$, otherwise $\eta_t^\ast=0$.  Since $w>0$,  $G$ always has a positive probability mass at 0 which could induce certain level of sparsity. In the following subsection, we'll develop a variational Algorithm to estimate $G$.

\subsection{Variational Algorithm for Estimating Prior}

Once we have specified Dirichlet Process as the prior, we need to compute the posterior distribution of $G$ in order to calculate Maximum A Posteriori (MAP) estimator of $G$. The major challenge here is that we have the infinite sum in the stick breaking form of $G$, which makes it impossible to sample from $G$. One remedy here is to fix $T$ as the upper bound of the number of clusters \citep{blei2006variational}. Then we have the following truncated version of stick breaking process using $G_0=w_0\delta_0+(1-w_0)\textsf{N}(0,\sigma_0^2)$ as the base measure.
\begin{align}\label{dp}
&V_t|\alpha_0 \sim \textsf{Beta}(1,\alpha_0), t=1,2,\cdots, T-1, V_T=1;\\
&\xi_t\sim \textsf{Ber}(w_0), t=1,2,\cdots, T;\\
&\eta^\ast_t|\xi_t \sim \left\{\begin{aligned}
&\delta_0 & \xi_t=1\\
&\textsf{N}(0,\sigma_0^2) & \xi_t=0\\
\end{aligned}\right.; t=1,2,\cdots,T;\\
&\pi_t =V_t\prod^{t-1}_{j=1}(1-V_j), t=1,2,\cdots, T-1, \pi_T =\prod^{T}_{j=1}(1-V_j);\\
&Z_k|\{V_1, V_2, \cdots, V_{T-1}\} \sim \text{Multinomial}(\boldsymbol{\pi});\\
&X_i |Z_i \sim \textsf{N}(\eta^\ast_{Z_i},1), i=1,2,\cdots, n.
\end{align}

The observed data are $\mathbf{y}$ and the parameters are $\mathbf{Z}_{1\times n},\mathbf{V}_{1\times (T-1)},\boldsymbol{\eta}^\ast_{1\times T},\boldsymbol{\xi}_{1\times T}$. $\boldsymbol{\eta}^\ast=(\eta_1^\ast,\cdots,\eta_T^\ast)$ contains all unique values of $\boldsymbol\eta=(\eta^\ast_{Z_i})^n_{i=1}$.

We could use Markov Chain Monte Carlo (MCMC) \citep{escobar1995bayesian} to compute the posterior distribution of these parameters. However, due to large $n$, the number of parameters is huge, making MCMC converging very slowly. In this paper, we use a variational algorithm to get posterior distribution. Variational Bayes methods are deterministic. The essence of variational inference is to regard the computation of posterior distribution as an optimization problem. Solving this optimization problem gives an approximation to the posterior distribution. \cite{blei2006variational} proposed variational algorithms for Dirichlet process mixture models for exponential family. Although normal distribution with a positive mass at 0 does not belong to exponential family, we could follow the same philosophy to deduce the corresponding algorithm.

We consider mean field variational inference and assume the following fully factorized variational distribution:
\[q(\mathbf{Z},\mathbf{V},\boldsymbol{\eta},\boldsymbol{\xi})=q_{\mathbf{p},\mathbf{m},\boldsymbol{\tau}}(\boldsymbol{\eta},\boldsymbol{\xi})q_{\boldsymbol{\gamma}_1,\boldsymbol{\gamma}_2}(\mathbf{V})q_{\boldsymbol{\Phi}}(\mathbf{Z});\]

Through the calculation shown in the Appendix, we could prove the optimal $q$ must be further factorized as follows:

\begin{itemize}
\item
$q_{\mathbf{p},\mathbf{m},\boldsymbol{\tau}}(\boldsymbol{\eta}^\ast,\boldsymbol{\xi})=\prod^T_{t=1}q_{p_t,m_t,\tau_t}(\eta^\ast_t,\xi_t)$,
where $\mathbf{p}=(p_1,p_2,\cdots,p_T)$, $\mathbf{m}=(m_1,m_2,\cdots,m_T)$, $\boldsymbol{\tau}=(\tau_1,\tau_2,\cdots,\tau_T)$, and $q_{p_t,m_t,\tau_t}(\eta^\ast_t,\xi_t)=p_t1_{\xi_t=1}\delta_0+(1-p_t)1_{\xi_t=0}q_{m_t,\tau_t}(\eta^\ast_t)$, where $q_{m_t,\tau_t}(\eta^\ast_t)$ is Normal density with mean $m_t$ and variance $\tau_t^2$.
\item
$q_{\boldsymbol{\gamma}_1,\boldsymbol{\gamma}_2}(\mathbf{V})=\prod^{T-1}_{t=1}q_{\gamma_{1t},\gamma_{2t}}(V_t)$, where
$\boldsymbol{\gamma}_1=(\gamma_{11},\gamma_{12},\cdots,\gamma_{1(T-1)})$, $\boldsymbol{\gamma}_2=(\gamma_{21},\gamma_{22},\cdots,\\ \gamma_{2(T-1)})$,
$q_{\gamma_{1t},\gamma_{2t}}(V_t)$ is Beta Distribution with parameters $(\gamma_{1t},\gamma_{2t})$.
\item
$q_{\boldsymbol{\Phi}}(\mathbf{Z})=\prod^n_{i=1}q_{\boldsymbol{\phi}_i}(Z_i)$;
where $\mathbf{Z}=(Z_1,Z_2,\cdots,Z_n)$, $\boldsymbol{\Phi}=(\boldsymbol{\phi}_1,\boldsymbol{\phi}_2,\cdots, \boldsymbol{\phi}_n)$,$\boldsymbol{\phi}_i=(\phi_{i,1},\phi_{i,2},\cdots, \phi_{i,T})$
, $\phi_{i,t}=q(Z_i=t)$, $q_{\boldsymbol{\phi}_i}(Z_i)$ is Multinomial distribution with parameters $\boldsymbol{\phi}_i$.
\end{itemize}

The algorithm is summarized in Algorithm 1. Via iterating these steps we could update the variational parameters. After convergence of $\boldsymbol{\Phi}$, $\mathbf{p}$, $\mathbf{m}$, $\boldsymbol{\tau}$, $\boldsymbol{\gamma}_1$ and $\boldsymbol{\gamma}_2$, we get an approximation of the posterior by plugging in these estimated parameters. The parameters we are interested in are $\boldsymbol{\Phi}$,$\mathbf{p}$ and $\mathbf{m}$. (In the algorithm $\|\cdot\|_{\infty,\infty}$ means the element-wise maximum absolute value; $\text{logit}(x)=(1+\exp(-x))^{-1}$.)

\begin{algorithm}[htbp]
\begin{algorithmic}
\State \textbf{input }{$\mathbf{y},\alpha_0,\sigma_0,w_0,T$}
\State \textbf{initialize} $\boldsymbol\Phi^{(1)}$ and $\boldsymbol\Phi^{(0)}$;
\While{$\|\boldsymbol\Phi^{(1)}-\boldsymbol\Phi^{(0)}\|_{\infty,\infty}>\epsilon$}
\State  $m_t\leftarrow\frac{\sigma_0^2\cdot \sum^n_{i=1}\phi^{(0)}_{i,t}X_i}{\sigma_0^2\cdot \sum^n_{i=1}\phi^{(0)}_{i,t}+1}, t=1,2,\cdots, T$;
\State   $\tau_t^2 \leftarrow\frac{\sigma_0^2}{\sigma_0^2\cdot \sum^n_{i=1}\phi^{(0)}_{i,t}+1}, t=1,2,\cdots,T$;
\State   $p_t\leftarrow\text{logit}^{-1}(\log(w_0)-\log(1-w_0)+\log(\sigma_0^2\cdot \sum^n_{i=1}\phi^{(0)}_{i,t}+1)/2-\frac{\sigma_0^2\cdot (\sum^n_{i=1}\phi^{(0)}_{i,t}X_i)^2}{2(\sigma_0^2\cdot \sum^n_{i=1}\phi^{(0)}_{i,t}+1)}),t=1,2,\cdots,T$;
\State   $\gamma_{t,1}\leftarrow 1+\sum^{n}_{i=1}\phi^{(0)}_{i,t}, t=1,2,\cdots, T-1$;
 \State $\gamma_{t,2}\leftarrow\alpha_0+\sum^{n}_{i=1}\sum^{T}_{j=t+1} \phi^{(0)}_{i,t}, t=1,2,\cdots, T-1$;
\State   $S_{i,t}\leftarrow\mathds{E}_{q}\log V_t+ \sum^{t-1}_{i=1}\mathds{E}_{q}\log (1-V_t)+ (1-p_t)m_{t}X_t-\frac{1}{2}(1-p_t)(m_{t}^2+\tau^2_t),t=1,2,\cdots, T, i=1,2,\cdots, n$;
 \State  $\phi^{(1)}_{i,t}\propto \exp(S_{i,t}), t=1,2,\cdots, T, i=1,2,\cdots, n$;
 \EndWhile
 \State \textbf{output } {$\mathbf{p},\mathbf{m},\boldsymbol{\Phi}$}
\end{algorithmic}
 \vspace{3mm}
 \caption{Variational Bayes Algorithm for Dirichlet process mixture model with $G_0$}
\end{algorithm}

\subsection{Posterior Computation}
Given approximate posterior estimates $\hat{\mathbf{p}}$, $\hat{\mathbf{m}}$, $\hat{\boldsymbol\Phi}$, we construct a MAP estimator of $G$. Remind that $\hat{m}_t$ is the nonzero cluster center; $\hat{p}_t$ is the probability mass of zero of component indexed by $t$; each entry $\hat{\phi}_{it}$ of $\hat{\boldsymbol{\Phi}}$ is the posterior probability of $Z_i$ belonging to the cluster $t$. The approximate posterior distribution of $\eta^\ast_{Z_i}$ is
\[(\sum^{T}_{t=1}\hat{\phi}_{it}\hat{p}_t)\delta_0(\cdot)+\sum^T_{t=1}\hat{\phi}_{it}(1-\hat{p}_t)\delta_{m_t}(\dot).\]
The most probable posterior assignment of $\eta^\ast_{Z_i}$ based on above posterior is denoted as $\hat{\eta}^\ast_{Z_i}$. MAP estimate of cluster weights including zero clusters is $\tilde{w}_t=\#\{k:\hat{\eta}^\ast_{Z_i}=\hat{m}_t\}/n$ and $\tilde{w}_0=\#\{k:\hat{\eta}^\ast_{Z_i}=0\}/n$. Then the estimated prior is
\[\hat{G}_{\text{MAP}}= \tilde{w}_0\cdot\delta_0(\cdot)+\sum^{T}_{t=1}\tilde{w}_t\delta_{\hat{m}_t}(\cdot).\]


Plugging in $\hat{G}_{\text{MAP}}$, data generation process we are considering could be written as ($\textsf{N}^{\kappa}$ denotes normal likelihood to the power of $\kappa$):
\begin{align*}
\theta_i&\sim \hat{G}_{\text{MAP}}, \hat{G}_{\text{MAP}}= \tilde{w}_0\cdot\delta_0(\cdot)+\sum^{T}_{t=1}\tilde{w}_t\delta_{\hat{m}_t}(\cdot)\\
X_i&\sim \textsf{N}^{\kappa}(\theta_i,1), i=1,2,\cdots, n.
\end{align*}

This is different from the original data generating process \ref{prior:new} but could be regarded as an approximation of \ref{prior:new} by setting $\sigma^2 = 0$.  The reason why we use this approximation is that instead of using MCMC, we could compute a closed-form posterior distribution of $\bth$ given $X^n$ using Bayes Formula: since $X_i\sim \textsf{N}(\theta_i,1)$ and $\theta_i\sim \hat{G}_{\text{MAP}}$, the posterior distribution of $\theta_i$ given $X^n$ has the form
\[\frac{\tilde{w}_0\exp(-\frac{\kappa X_i^2}{2})\delta_0+\sum^T_{t=1}\tilde{w}_t\exp(-\frac{\kappa(X_i-\hat{m}_t)^2}{2})\delta_{\hat{m}_t}}{\tilde{w}_0\exp(-\frac{\kappa X_i^2}{2})+\sum^T_{t=1}\tilde{w}_t\exp(-\frac{\kappa(X_i-\hat{m}_t)^2}{2})}\equiv \hat{w}_0\cdot\delta_0(\cdot)+\sum^{T}_{t=1}\hat{w}_t\delta_{\hat{m}_t}(\cdot);\]
where $\hat{w}_t$ is the posterior weight and $\sum^T_{t=1}\hat{w}_t=1$. The final estimate of $\theta_i$ is the posterior mean estimator
\[\hat{\theta}_i=\sum^{T}_{t=1}\hat{w}_t\hat{m}_t.\]

\subsection{Simulation Studies}

We conduct the following four simulation studies; R package for our empirical Bayes estimator is available in \url{https://github.com/yunboouyang/VBDP}; all the source code is summarized in \url{https://github.com/yunboouyang/EBestimator}. For our empirical Bayes estimator, we set the maximal number of clusters to be $T=10$. We set fractional likelihood parameter $\kappa=0.99$ throughout all the simulation study. We set concentration parameter $\alpha_0=1$ in all of the simulation studies. For $G_0=w_0\delta_{0}+(1-w_0)\textsf{N}(0,\sigma_0^2)$, we set $w_0=0.01$. For simulation study 1, we set $\sigma_0=4$. For other simulation studies, we set $\sigma_0=6$. The influence of these hyper-parameters will diminish when sample size $n$ is large.

Besides our method (denoted as DP representing Dirichlet process clustering), we also consider estimators proposed by \cite{martin:walker:2014} (denoted as EBMW), by  \cite{koenker:mizera:2014} (denoted as EBKM), and by \cite{johnstone2004needles}(denoted as EBMed, we directly use functions from the \texttt{EbayesThresh} package by \cite{Johnstone:Silverman:NeedleComput}), as well as the Hard thresholding estimator, soft thresholding estimator, SURE estimator (using \texttt{waveThresh} package) and FDR estimator with parameters $q=0.01,0.1,4$.

\subsubsection*{Experiment 1}
In the first simulation study, we take sample $X^n$ of dimension $n=200$ from the normal mean model $X_i\sim \textsf{N}(\theta_i,1)$. In this case, we consider the number of nonzero elements to be $s_n=10, 20, 40, 80$ and the signals are fixed at values $\mu_0=1,3,5,7$. We compute mean squared error (MSE) and mean absolute error (MAE) based on 200 replications as two measure of performance. We summarize the results in Table 1 and Table 2.
\begin{table}[H]
\footnotesize
\begin{center}
\begin{tabular}{ccccccccccccccccc}
\hline
$s_n$ & \multicolumn{4}{c}{10} &  \multicolumn{4}{c}{20} & \multicolumn{4}{c}{40} & \multicolumn{4}{c}{80}\\
$\mu_0$ &1 &3 &5 &7 &1 &3 &5 &7 &1 &3 &5 &7 &1 &3 &5 &7 \\
\hline
EBMW & \textbf{10} &  54 &  21 &  13 &  20 &  96 &  35  & 25 &  40 &  152 &   61 &   49 &   79  & 234  & 108  &  96\\
EBKM & 12 &  \textbf{35} &  15  &  6  & 21 &  53 &  19 &   7 &  \textbf{32}  &  74  &  25  &   7 &   \textbf{44}  &  93  &  30  &   8\\
EBmed & \textbf{10} &  43  & 22 &  14  & 20 &  69  & 36  & 28 &  37 &  103 &   66   & 54  &  64 &  157  & 127 &  107\\
SURE & 14  & 42 &  45 &  43 &  23 &  68 &  69  & 69  & 42 &  104 &  105 &  105   & 74  & 152  & 153  & 153\\
Soft Thresholding &\textbf{10} &  76  &116 & 116  & 20 & 153&  228 & 233 &  40 &  304 &  457  & 464  &  80  & 609  & 916 &  929\\
Hard Thresholding  & 13 &  61 &  21  & 12  & 24 & 122 &  39 &  23 &  45  & 237  &  75 &   42  &  87  & 473 &  149 &   82\\
FDR $q=0.01$ & \textbf{10}  & 75  & 25 &  11 &  20 & 143 &  40 &  22 &  41 &  253  &  67  &  44  &  81 &  434 &  112  &  85\\
FDR $q=0.1$  & 11 &  55 &  24 &  20 &  23 &  93 &  39  & 37 &  44 &  141  &  67   & 64  &  88 &  208 &  113 &  111\\
FDR $q=0.4$  & 22 &  63 &  53 &  51 &  36&   94  & 82 &  80  & 64 &  134  & 119  & 117  & 119  & 175  & 161  & 161\\
\hline
DP  &11 &37 &\textbf{11} & \textbf{3} &\textbf{19} &\textbf{50} &\textbf{17} & \textbf{4}& 33 &\textbf{71} &\textbf{22} & \textbf{4} &46 &\textbf{92}& \textbf{26} & \textbf{6}\\
\hline
\end{tabular}
\end{center}
\caption{MSE of Simulation Study 1, error of the best method marked as bold}
\end{table}

\begin{table}[H]
\footnotesize
\begin{center}
\begin{tabular}{ccccccccccccccccc}
\hline
$s_n$ & \multicolumn{4}{c}{10} &  \multicolumn{4}{c}{20} & \multicolumn{4}{c}{40} & \multicolumn{4}{c}{80}\\
$\mu_0$ &1 &3 &5 &7 &1 &3 &5 &7 &1 &3 &5 &7 &1 &3 &5 &7 \\
\hline
EBMW & \textbf{10} &  23  & 14  & 13 &  \textbf{20} &  43  & 26  & 24 &  \textbf{40} &   74  &  47 &   45  &  \textbf{80} &  125  &  86 &   83\\
EBKM & 27  & 35  & 22  & 19  & 40  & 48 &  25 &  20 &  60  &  62  &  29  &  21  &  82   & 76  &  32  &  22\\
EBmed & 12 &  \textbf{21} &  12  & 10 &  23 &  38 &  23 &  20 &  44  &  72  &  45 &   39  &  \textbf{80} &  133  &  96  &  80\\
SURE & 15  & 30  & 32  & 31 &  24  & 49 &  51 &  51  & 44  &  78  &  79  &  79  &  78  & 117  & 119  & 119\\
Soft Thresholding  & \textbf{10} &  27 &  33 &  33 &  20 &  54 &  65 &  65  & \textbf{40}  & 109  & 130 &  130  &  \textbf{80} &  217  & 260 &  261\\
Hard Thresholding & 11  & 22  & \textbf{10} &   9 &  21 &  45  & 19 &  17  & 41  &  88  &  37  &  32  &  82 &  175  &  74   & 64 \\
FDR $q=0.01$ & \textbf{10}  & 27  & \textbf{10}  &  \textbf{8}  & \textbf{20} &  51 &  19  & 17  & \textbf{40}  &  93  &  36  &  33  &  \textbf{80}  & 163 &   69  &  66\\
FDR $q=0.1$ &  \textbf{10} &  \textbf{21}  & 12 &  11  & 21 &  \textbf{36} &  22 &  22 &  41  &  59 &   41 &   41  &  82  &  98  &  77  &  76\\
FDR $q=0.4$ & 14 &  26 &  25 &  25 &  25 &  44  & 43  & 42 &  48  &  72  &  70 &   70  &  94  & 110 &  108 &  108\\
\hline
DP& 23 &31 &18 &14& 36 &42 &\textbf{20} &\textbf{16} &61 &\textbf{57} &\textbf{25} &\textbf{17} &87 &\textbf{72} &\textbf{27} &\textbf{19}\\
\hline
\end{tabular}
\end{center}
\caption{MAE of Simulation Study 1, error of the best method marked as bold}
\end{table}

When $\mu_0$ is larger than one, our DP method is the best one in terms of MSE since it is easier to figure out the cluster centers of $\theta_i$'s and estimate the nonzero $\theta_i$'s. Besides, when $s_n$ is large, our DP method has the best performance among all since the more nonzero $\theta_i$'s we have, the easier we could estimate the cluster centers. When both $\mu_0$ and $\theta_i$ are small, which is a tough case since high dimensional noise might make it much more difficult to detect weak signals, our method still has the comparable performance among all methods. Our method has better performance when measured by MSE than that in MAE. Other methods such as EBKM and EBmed also have comparable good performance.

\subsubsection*{Experiment 2}
In the second simulation study we increase the data dimension to 500. We are interested in the case when $s_n=25, 50, 100$ and all nonzero elements are fixed at $\mu_0=3,4,5$. This is a more challenging case since the signal is not very strong. MSE and MAE are shown in Table 3 and Table 4 respectively.

\begin{table}[H]
\small
\begin{center}
\begin{tabular}{cccccccccc}
\hline
$s_n$ & \multicolumn{3}{c}{25} &  \multicolumn{3}{c}{50} & \multicolumn{3}{c}{100}\\
$\mu_0$  &3 &4 &5  &3 &4 &5  &3 &4 &5\\
\hline
EBMW & 138  & 99  & 52 & 233 & 156 &  90 & 385 & 248 & 153 \\
EBKM & 81 &  57 &  28 & 120  & 80 &  41 & 174 & 114 &  52 \\
EBMed &  107 &  78 &  50 & 165 & 124 &  91 & 254 & 208 & 163\\
SURE & 102 & 106 & 105 & 163 & 168 & 170 & 257 & 260 & 259\\
ST &201 & 289 & 327 & 403 & 577 & 658 & 806 &1155 & 1308\\
HT &  172 & 143 &  64 & 341 & 282 & 129 & 680 & 563 & 251\\
FDR $q=0.01$ &  192 & 151 &  62 & 356 & 241 & 101 & 639 & 385 & 164\\
FDR $q=0.1$ & 139 &  87 &  55 & 224 & 135 &  99 & 355 & 213 & 167\\
FDR $q=0.4$ &  150 & 133 & 126 & 229 & 206 & 201 & 332 & 302 & 294\\
\hline
DP  & \textbf{80} & \textbf{55} & \textbf{25} &\textbf{119} & \textbf{79} & \textbf{35} &\textbf{171} &\textbf{109} & \textbf{49}\\
\hline
\end{tabular}
\end{center}
\caption{MSE of Simulation Study 2, error of the best method marked as bold}
\end{table}

\begin{table}[H]
\small
\begin{center}
\begin{tabular}{cccccccccc}
\hline
$s_n$ & \multicolumn{3}{c}{25} &  \multicolumn{3}{c}{50} & \multicolumn{3}{c}{100}\\
$\mu_0$  &3 &4 &5  &3 &4 &5  &3 &4 &5\\
\hline
EBMW &  58 &  47  & 36 & 105 &  82 &  66 & 186 & 143 & 118\\
EBKM &  75  & 54  & 38 & 102 &  67  & 45 & 139 &  84 &  50\\
EBMed & \textbf{49} & \textbf{37}  & 29 &  88 &  69  & 58 & 176 & 135 & 113\\
SURE &  73 &  76  & 75 & 120 & 124 & 124 & 194 & 196 & 196\\
ST & 70 &  83 &  87 & 141 & 166 & 175 & 281 & 332 & 349\\
HT &   62  & 44 &  \textbf{26} & 123 &  86 &  52 & 245 & 172 & 102\\
FDR $q=0.01$ &  67 &  46 &  \textbf{26} & 127 &  77 &  48 & 233 & 134 &  90 \\
FDR $q=0.1$ &  52  & 34 &  29  & \textbf{87} &  61 &  56 & 148 & 111 & 103\\
FDR $q=0.4$ &   63  & 61 &  60 & 107 & 106 & 106 & 179 & 175 & 174\\
\hline
DP &60 & 42 & 29 & 93 & \textbf{58} & \textbf{34} &\textbf{128} & \textbf{74} & \textbf{43}\\
\hline
\end{tabular}
\end{center}
\caption{MAE of Simulation Study 2, error of the best method marked as bold}
\end{table}

In terms of MSE, our DP method is consistently the best one among all the different configurations. The performance of EBKM is comparable. One drawback of EBKM is that the computational cost is high since solving a high dimensional optimization problem is needed. However for DP, we use a computationally efficient variational  inference algorithm, which dramatically saves the computational time.

\subsubsection*{Experiment 3}
In the third simulation study we maintain all the features of the second experiment except
that non-zero elements of the true mean vector are now generated from standard Gaussian distribution centered at the original values. All different $\theta_i$'s form a data cloud which is centered at the distinct non-zero values in simulation study 2.

\begin{table}[H]
\small
\begin{center}
\begin{tabular}{cccccccccc}
\hline
$s_n$ & \multicolumn{3}{c}{25} &  \multicolumn{3}{c}{50} & \multicolumn{3}{c}{100}\\
$\mu_0$  &3 &4 &5  &3 &4 &5  &3 &4 &5\\
\hline
EBMW &  110 &  91 &  62 & 186 & 150 & 103 & 308 & 245&  176\\
EBKM &  77  & \textbf{68}  &\textbf{ 49} & 121 & \textbf{105} &  \textbf{77} & \textbf{185} & \textbf{163} & 124\\
EBMed &  93  & 79  & 59 & 151 & 130 & 100 & 236 & 212 & 176\\
SURE &  96 & 104 & 106&  155 & 165 & 168&  243 & 257 & 260\\
ST & 196 & 269 & 313 & 389 & 534 & 626 & 780& 1075 &1251 \\
HT &   135 & 123 &  79 & 263 & 237 & 155 & 527 & 479 & 311\\
FDR $q=0.01$ &  151 & 129 &  79 & 268 & 216 & 133 & 479 & 368 & 221\\
FDR $q=0.1$ & 112 &  90  & 66 & 187 & 148 & 110 & 300 & 235 & 187\\
FDR $q=0.4$ &   138 & 137 & 129 & 216 & 213 & 201 & 320 & 307 & 301\\
\hline
DP &    \textbf{76} & \textbf{68} & 53 &\textbf{120} &110 & 84 &189 &164 &\textbf{123}\\
\hline
\end{tabular}
\end{center}
\caption{MSE of Simulation Study 3, error of the best method marked as bold}
\end{table}

\begin{table}[H]
\small
\begin{center}
\begin{tabular}{cccccccccc}
\hline
$s_n$ & \multicolumn{3}{c}{25} &  \multicolumn{3}{c}{50} & \multicolumn{3}{c}{100}\\
$\mu_0$  &3 &4 &5  &3 &4 &5  &3 &4 &5\\
\hline
EBMW &  51  & 46  & 38  & 92  & 82  & 69 & 164 & 145 & 126\\
EBKM &   75 &  63  & 52 & 108  & 91  & 74 & 160 & 135 & 111\\
EBMed &   \textbf{44} &  38 &  \textbf{31} &  \textbf{79} &  \textbf{70} &  59 & 147 & 132 & 115\\
SURE &   69 &  74 &  76 & 114 & 122 & 124 & 187 & 195 & 197\\
ST &   67  & 79 &  86 & 133 & 158 & 171 & 266 & 317 & 342\\
HT &   51  & 43  & \textbf{31} & 100 &  84  & 61 & 200 & 169 & 121 \\
FDR $q=0.01$ &    55 &  44 &  \textbf{31 }& 101 &  80 &  \textbf{57} & 188 & 144 & 104\\
FDR $q=0.1$ &   45  & \textbf{37} & \textbf{31} &  81 &  67 &  58 & \textbf{143} & \textbf{120} & 108\\
FDR $q=0.4$ &   60  & 62 &  60 & 105 & 107 & 105 & 176 & 175 & 176\\
\hline
DP &    59 & 54 & 44 & 97 & 84 & 69 &156 &127 &\textbf{102}\\
\hline
\end{tabular}
\end{center}
\caption{MAE of Simulation Study 3, error of the best method marked as bold}
\end{table}

Even though our DP method is not as good as EBKM and EBMed in some configurations, our DP method has quite similar performance even it's more difficult to estimate the clustering centers in Simulation Study 3 than Simulation Study 2.

\subsubsection*{Experiment 4}
In the fourth example, we consider $1000$-dimensional vector estimation, with the first 10 entries of $\bth^\ast$ equal 10, the next 90 entries equal $A$, and the remaining 900 entries equal 0. We consider a range of $A$, from $A=2$ to $A=7$. DP method will automatically estimate 3 cluster centers instead of 2. Average MSE and MAE are recorded in Table 7 and Table 8.

\begin{table}[H]
\begin{center}
\begin{tabular}{ccccccc}
\hline
$A$ & 2& 3& 4&  5& 6& 7\\
\hline
EBMW &  320 & 421 & 291 & 175 & 134 & 125\\
EBKM &  206 & 223 & \textbf{150}  & \textbf{79}  & \textbf{46}  & \textbf{35}\\
EBMed & 337 & 350 & 240 & 173 & 148 & 138 \\
SURE & 278 & 327 & 333 & 337 & 336 & 334 \\
ST &   504 & 894 &1259 &1436 &1475 &1483\\
HT & 375 & 671 & 610 & 301 & 134 & 102\\
FDR $q=0.01$ & 375 & 633 & 441 & 199 & 121 & 111 \\
FDR $q=0.1$ & 373 & 421 & 259 & 194 & 185 & 181 \\
FDR $q=0.4$ &440 & 451 & 404 & 403 & 399 & 396 \\
\hline
DP &  \textbf{204} &\textbf{220} &161 &205 &151 & 85\\
\hline
\end{tabular}
\end{center}
\caption{MSE of Simulation 4, error of the best method is marked as bold}
\end{table}

\begin{table}[H]
\begin{center}
\begin{tabular}{ccccccc}
\hline
$A$ & 2& 3& 4&  5& 6& 7\\
\hline
EBMW & 179 & 199 & 159 & 130 & 122 & 120\\
EBKM &225 & 182 & 115  & \textbf{79}  & \textbf{61} &  61 \\
EBMed &  \textbf{176} & \textbf{161} & 127 & 110 & 101 &  97\\
SURE & 209 & 241 & 245 & 248 & 246 & 247 \\
ST &  216 & 294 & 347&  367 & 371 & 372\\
HT & 189 & 242 & 184 & 110 &  85  & 80\\
FDR $q=0.01$ & 189 & 232 & 147 &  96 &  85 &  83 \\
FDR $q=0.1$ &  188 & 168 & 120 & 110 & 109 & 108\\
FDR $q=0.4$ & 217 & 214 & 208 & 212 & 210 & 209 \\
\hline
DP &   215 &\textbf{161} & \textbf{91} & 89 & 75 & \textbf{57}\\
\hline
\end{tabular}
\end{center}
\caption{MAE of Simulation 4,  error of the best method marked as bold}
\end{table}

In some configurations, DP method is the best one among all the state-of-the-art methods. Although in some cases EBKM and EBMed performs better, MSE and MAE of our DP method are still comparable.

\section{Conclusions and Discussions}

Empirical Bayes method introduced in \cite{martin:walker:2014} does not have good performance in simulation study since a prior centered at $X_i$ may introduce noise in posterior estimation, whereas our nonparametric Bayes based clustering method will decrease the noise but capture the general pattern: the estimate of $\theta_i$ is adaptively shrunk toward the mean of nearby data points. Therefore our DP method outperforms the empirical Bayes estimator based on \cite{martin:walker:2014}.

Theoretical results shows with proper choice of parameters in the prior, our posterior mean estimator achieves asymptotically minimax rate. In the implementation we propose a fast variational inference method which approximates posterior distribution, which is more efficient than Empirical Bayes method proposed by \cite{koenker:bakeoff:2014} in terms of computation time. To our knowledge, this is a first work connecting high dimensional sparse vector estimation with clustering. Our nonparametric Bayesian estimator could be applied to high dimensional classification, feature selection, hypothesis testing and nonparametric function estimation. Possible extension in implementation includes using other well-studied clustering methods to estimate prior and comparing their performance.

\section*{Proofs}

\subsection*{Proof for Lemma \ref{lemma:denominator}}

Denote $\mathbf{w}=(w_1,w_2,\cdots, w_T)$ and $\mathbf{m}=(m_1,m_2,\cdots, m_T)$. Write $D_n$ in terms of the conditional prior $(\theta_1,\theta_2,\cdots, \theta_n)| w \sim \Pi_{w, \mathbf{w}, \mathbf{m}}$ and the marginal prior $w\sim \Pi$. Therefore

\[D_n=\int^1_0 \prod^{n}_{i=1}\int_\RR\{\frac{p_{\theta_i}(X_i)}{p_{\theta_i^\ast}(X_i)}\}^{\kappa}\Pi_{w,\mathbf{w}, \mathbf{m}}(d\theta_i)\Pi(dw);\]
where $p_{\theta_i}(X_i)=\frac{1}{\sqrt{2\pi}}\exp(-\frac{(X_i-\theta_i)^2}{2})$.  For given $w$, the inner expectation involves an average over all configurations of the indicators $(1_{\theta_1=0},1_{\theta_2=0},\cdots, 1_{\theta_n=0})$. This average is clearly larger than just the case where the indicators exactly match up with the support $S^\ast$ of $\theta^\ast$, times the probability of that configuration, that is,
\[D_n>\int^1_0 w^{n-s_n}(1-w)^{s_n}\Pi(dw)\prod_{i\in S^\ast}\int_\RR e^{\frac{\kappa}{2}\{(X_i-\theta^\ast_i)^2-(X_i-\theta_i)^2\}}\sum^T_{t=1}\frac{w_t}{\sqrt{2\pi\sigma^2}}e^{-\frac{1}{2\sigma^2}(m_t-\theta_i)^2}d\theta_i. \]

For each $i\in S^\ast$ and each $1\leq t\leq T$,
\begin{eqnarray*}
&& \int_\RR e^{\frac{\kappa}{2}\{(X_i-\theta^\ast_i)^2-(X_i-\theta_i)^2\}}\frac{w_t}{\sqrt{2\pi\sigma^2}}e^{-\frac{1}{2\sigma^2}(m_t-\theta_i)^2}d\theta_i \\
&=& e^{\frac{\kappa}{2}(X_i-\theta^\ast_i)^2}\int_\RR e^{-\frac{\kappa}{2}(X_i-\theta_i)^2}\frac{w_t}{\sqrt{2\pi\sigma^2}}e^{-\frac{1}{2\sigma^2}(m_t-\theta_i)^2}d\theta_i\\
&=& e^{\frac{\kappa}{2}(X_i-\theta^\ast_i)^2}\frac{w_t}{\sqrt{\kappa\sigma^2+1}}\exp(-\frac{\kappa(X_i-m_t)^2}{\kappa\sigma^2+1}).
\end{eqnarray*}
Therefore we have
\begin{eqnarray*}
&& \prod_{i\in S^\ast}\int_\RR e^{\frac{\kappa}{2}\{(X_i-\theta^\ast_i)^2-(X_i-\theta_i)^2\}}\sum^T_{t=1}\frac{w_t}{\sqrt{2\pi\sigma^2}}e^{-\frac{1}{2\sigma^2}(m_t-\theta_i)^2}d\theta_i\\
&=& \prod_{i\in S^\ast}\{e^{\frac{\kappa}{2}(X_i-\theta^\ast_i)^2}\sum^T_{t=1}\frac{w_t}{\sqrt{\kappa\sigma^2+1}}\exp(-\frac{\kappa(X_i-m_t)^2}{\kappa\sigma^2+1})\}\\
&=& (\kappa\sigma^2+1)^{-\frac{s_n}{2}}\prod_{i\in S^\ast}\{e^{\frac{\kappa}{2}(X_i-\theta^\ast_i)^2}\}\prod_{i\in S^\ast}\{\sum^T_{t=1}w_t\exp(-\frac{\kappa(X_i-m_t)^2}{\kappa\sigma^2+1})\}.
\end{eqnarray*}
Since
\begin{eqnarray*}
&& \prod_{i\in S^\ast}\{\sum^T_{t=1}w_t\exp(-\frac{\kappa(X_i-m_t)^2}{\kappa\sigma^2+1})\}\geq  \prod_{i\in S^\ast}\{w_{t_i}\exp(-\sum^T_{t=1}\frac{\kappa (X_i-m_{t_i})^2}{\kappa\sigma^2+1})\}\\
&=&(\prod_{i\in S^\ast}w_{t_i})\times \exp(-\kappa\frac{\sum_{i\in S^\ast} (X_i-m_{t_i})^2}{\kappa\sigma^2+1}).
\end{eqnarray*}
We have $\prod_{i\in S^\ast}w_{t_i}\geq (Cs_n/n)^{s_n}=\exp(\log C \cdot s_n-\varepsilon_n)$. Since $\sum_{i \in S^\ast} (X_i-m_{t_i})^2\leq 2\sum_{i \in S^\ast} (\theta_i^\ast-m_{t_i})^2 + 2\sum_{i \in S^\ast} (X_i-\theta_i^\ast)^2 $ and $\sum_{i \in S^\ast} (X_i-\theta_i^\ast)^2\sim \chi^2_{s_n}$, by concentration inequality, we have
\[P(|\sum_{i \in S^\ast} (X_i-\theta_i^\ast)^2/s_n-1|\geq \log(n/s_n)-1)\leq 2\exp(-s_n(\log(n/s_n)-1)^2/8).\]
Therefore $\sum_{i \in S^\ast}(X_i-\theta_i^\ast)^2=o_p(\varepsilon_n)$. Hence
\[\prod_{i\in S^\ast}\{\sum^T_{t=1}w_t\exp(-\frac{\kappa(X_i-m_t)^2}{\kappa\sigma^2+1})\}\geq \exp(\log C \cdot s_n-\varepsilon_n-\frac{\kappa}{\kappa\sigma^2+1}o_p(\varepsilon_n)).\]

According to Law of Large Numbers,
\[\prod_{i\in S^\ast}e^{\frac{\kappa}{2}\{(X_i-\theta^\ast_i)^2\}}=e^{\frac{\kappa}{2}\prod_{i\in S^\ast}\{(X_i-\theta^\ast_i)^2\}}=e^{\frac{\kappa s_n}{2}+o_p(s_n)}. \]

Therefore
\[\prod_{i\in S^\ast}\int_\RR e^{\frac{\kappa}{2}\{(X_i-\theta^\ast_i)^2-(X_i-\theta_i)^2\}}\sum^T_{t=1}\frac{w_t}{\sqrt{2\pi\sigma^2}}e^{-\frac{1}{2\sigma^2}(m_t-\theta_i)^2}d\theta_i\geq(1+\kappa\sigma^2)^{-\frac{s_n}{2}}\exp(-\varepsilon_n-o_p(\varepsilon_n)). \]
Using the same trick in the proof of Lemma 1 in \cite{martin:walker:2014},  $\int^{1}_{0}w^{n-s_n}(1-w)^{s_n}\Pi(dw)$ is lower bounded by $\frac{\alpha}{1+\alpha}\exp[-2\varepsilon_n-\alpha s_n+o(s_n)]$ as $n$ is sufficiently large. Putting these pieces together, we obtain
\[D_n \ge \frac{\alpha}{1+\alpha}\exp\{-3\varepsilon_n-o_p(\varepsilon_n)\}. \]

\subsection{Proof for Theorem \ref{thm:concentration}}

The main aim of the proof is to show the numerator, for sets $A_n$ away from $\bth^\ast$, is not too large. Let $N_n$ be the numerator for $Q_n(A_{M\varepsilon_n})$, i.e.,
\[N_n=\int^1_0~~\int_{A_{M\varepsilon_n}} \prod^{n}_{i=1} \big(\{\frac{p_{\theta_i}(X_i)}{p_{\theta_i^\ast}(X_i)}\}^{\kappa}\Pi_{w,\mathbf{w},\mathbf{m}}(d\theta_i)\big)~~\Pi(dw). \]
Taking expectation of $N_n$, with respect to $P_{\theta^\ast}$, we get
\[\EE_{\bth^\ast}(N_n)=\int^1_0~~\int_{A_{M\varepsilon_n}} ~\big(\prod^{n}_{i=1}\int_\RR\{\frac{p_{\theta_i}(x_i)}{p_{\theta_i^\ast}(x_i)}\}^{\kappa}p_{\theta^\ast_i}(x_i)dx_i\big)~\Pi_{w,\mathbf{w},\mathbf{m}}(d\theta_i)~~\Pi(dw).\]

Since $\int_\RR\{\frac{p_{\theta_i}(x_i)}{p_{\theta_i^\ast}(x_i)}\}^{\kappa}p_{\theta^\ast_i}(x_i)dx_i=\exp\{-\frac{\kappa(1-\kappa)}{2}(\theta_i-\theta_i^\ast)^2\}$, we have
\[\EE_{\bth^\ast}(N_n)=\int^1_0~~\int_{A_{M\varepsilon_n}} ~\prod^{n}_{i=1}\exp\{-\frac{\kappa(1-\kappa)}{2}(\theta_i-\theta_i^\ast)^2\}~\Pi_{w,\mathbf{w},\mathbf{m}}(d\theta_i)~~\Pi(dw).\]
Since $ A_{M\varepsilon_n}=\{\bth|~~\|\bth -\bth^\ast\|^2>M\varepsilon_n\}$, we have
\[\EE_{\bth^\ast}(N_n)\leq \exp\{-\frac{\kappa(1-\kappa)M}{2}\varepsilon_n\}.\]

Let $\frac{\kappa(1-\kappa)}{2}=c$. Next, take $M$ such that $cM>3$, and then take $K\in (3, cM)$, then using Markov Inequality we get $P_{\theta^\ast}(N_n>e^{-K\varepsilon_n})\leq e^{-(cM-K)\varepsilon_n}$. This upper bound has a finite sum over $n\geq 1$, so the Borel-Catelli lemma gives that $N_n\leq e^{-K\varepsilon_n}$ with probability 1 for all large $n$. Therefore we get

\[\frac{N_n}{D_n}\leq \frac{1+\alpha}{\alpha}e^{-(K-3)\varepsilon_n-\eta s_n+o(s_n)}. \]
Since $s_n=o(\varepsilon_n)$, $Q_n(A_{M\varepsilon_n})\rightarrow 0$ as $n\rightarrow \infty$.

\subsection*{Variational Inference Algorithm Derivation}
We will derive the variational inference algorithm for Dirichlet process mixture model. $\alpha_0$, $T$, $w_0$, $\sigma_0^2$ and the data vector $X^n$ is given in advance. The data generating process is summarized in \ref{dp}. We treat $\mathbf{Z}$ as latent variables and $\mathbf{V},\boldsymbol{\eta^\ast},\boldsymbol{\xi}$ as parameters. Posterior distribution of all the parameters and latent variables is proportional to
\begin{align*}
P(\mathbf{Z},\mathbf{V},\boldsymbol{\eta^\ast},\boldsymbol{\xi}|X^n)& \propto P(X^n,\mathbf{Z},\mathbf{V},\boldsymbol{\eta^\ast},\boldsymbol{\xi})=P(\boldsymbol{\xi}|w)P(\mathbf{V}|\alpha_0)P(\boldsymbol{\eta^\ast}|\boldsymbol{\xi})P(\mathbf{Z}|\mathbf{V})P(X^n|\mathbf{Z},\boldsymbol{\eta^\ast})\\
&\propto w_0^{\sum^T_{t=1}\xi_t}(1-w_0)^{T-\sum^T_{t=1}\xi_t}\prod^{T-1}_{t=1}(1-V_t)^{\alpha_0-1}\prod_{t:\xi_t=1}\delta_0(\eta^\ast_t)\cdot\\
&\prod_{t:\xi_t=0}\frac{1}{\sqrt{2\pi}\sigma_0}\exp(-\frac{(\eta^\ast_t)^2}{2\sigma_0^2})\cdot\prod_{t=1}^T\pi_t^{\sum^n_{i=1}1_{Z_i=t}}\cdot\\
&\exp\left(-\frac{\sum^n_{i=1}\sum^T_{t=1}(X_i-\eta^\ast_t)^21_{Z_i=t}}{2}\right).
\end{align*}
Recall that under the fully factorized variational assumption, we have
\[q(\mathbf{Z},\mathbf{V},\boldsymbol{\eta^\ast},\boldsymbol{\xi})=q_{\mathbf{p},\mathbf{m},\boldsymbol{\tau}}(\boldsymbol{\eta^\ast},\boldsymbol{\xi})q_{\boldsymbol{\gamma}_1,\boldsymbol{\gamma}_2}(\mathbf{V})q_{\boldsymbol{\Phi}}(\mathbf{Z}).\]
Define $P(Z_i=t)=\phi_{i,t}$. First we find the optimal form of $q(\boldsymbol{\eta^\ast},\boldsymbol{\xi})$, which satisfies
\begin{align*}
\log q(\boldsymbol{\eta^\ast},\boldsymbol{\xi})&=\mathds{E}_{\mathbf{V},\mathbf{Z}}[\log(w_0^{\sum^T_{t=1}\xi_t}(1-w_0)^{T-\sum^T_{t=1}\xi_t}\prod_{t:\xi_t=1}\delta_0(\eta^\ast_t)\cdot\\
&\prod_{t:\xi_t=0}\frac{1}{\sqrt{2\pi}\sigma_0}\exp(-\frac{(\eta^\ast_t)^2}{2\sigma_0^2})\exp(-\frac{\sum^n_{i=1}\sum^T_{t=1}(X_i-\eta^\ast_t)^21_{Z_i=t}}{2}))]+\text{const}\\
&=\sum^T_{t=1}[1_{\xi_t=1}(\log w_0+\log \delta_0(\eta^\ast_t))+1_{\xi_t=0}(\log (1-w_0)-\log \sqrt{2\pi \sigma_0^2}-\frac{(\eta^\ast_t)^2}{2\sigma_0^2})\\
&-\frac{\sum^n_{i=1}\phi_{i,t}(X_i-\eta^\ast_t)^2}{2}]+\text{const}\\
&=\sum^T_{t=1}[1_{\xi_t=1}(\log w_0+\log \delta_0(\eta^\ast_t)-\frac{\sum^n_{i=1}\phi_{i,t}X_i^2}{2})\\
&+1_{\xi_t=0}(\log (1-w_0)-\log \sqrt{2\pi \sigma_0^2}-\frac{(\eta^\ast_t)^2}{2\sigma_0^2}-\frac{\sum^n_{i=1}\phi_{i,t}(X_i-\eta^\ast_t)^2}{2})+\text{const}]\\
&\equiv \sum^T_{t=1}\log q(\xi_t,\eta^\ast_t);
\end{align*}
where $\log q(\xi_t,\eta^\ast_t)=1_{\xi_t=1}(\log w_0+\log \delta_0(\eta^\ast_t)-\frac{\sum^n_{i=1}\phi_{i,t}X_i^2}{2})
+1_{\xi_t=0}(\log (1-w_0)-\log \sqrt{2\pi \sigma_0^2}-\frac{(\eta^\ast_t)^2}{2\sigma_0^2}-\frac{\sum^n_{i=1}\phi_{i,t}(X_i-\eta^\ast_t)^2}{2})+\text{const}$. Therefore the optimal form of $q_{\mathbf{p},\mathbf{m},\boldsymbol{\tau}}(\boldsymbol{\eta^\ast},\boldsymbol{\xi})$ is fully factorized across different clusters:
\[q_{\mathbf{p},\mathbf{m},\boldsymbol{\tau}}(\boldsymbol{\eta^\ast},\boldsymbol{\xi})=\prod^T_{t=1}q_{p_t,m_t,\tau_t}(\eta^\ast_t,\xi_t).\]
In order to determine the updating formula for $p_t,m_t,\tau_t$, we use Method of Undetermined Coefficients. Suppose $q(\xi_t,\eta^\ast_t)$ is a mixture of a point mass of zero and normal distribution,
 \[q(\xi_t,\eta^\ast_t)= p_t1_{\xi_t=1}\delta_0(\eta^\ast_t)+(1-p_t)1_{\xi_t=0} (2\pi\tau^2_t)^{-1/2}\exp(-(\eta^\ast_t-m_t)^2/(2\tau^2_t)),\]
therefore
\[\log q(\xi_t,\eta^\ast_t)=1_{\xi_t=1}(\log p_t+\log(\delta_0(\eta^\ast_t)))+1_{\xi_t=0}(\log(1-p_t)-\log(\sqrt{2\pi \tau_t^2})-\frac{(\eta^\ast_t-m_t)^2}{2\tau_t^2})+\text{const}.\]
Even though there's a normalizing constant, but the difference between multipliers of $1_{\xi_t=1}$ and $1_{\xi_t=0}$ is invariant with respect to the constant. Therefore we have the following equation:
\begin{align*}
&\log p_t-\log(1-p_t)+\log \sqrt{2\pi \tau_t^2}+\frac{(\eta^\ast_t-m_t)^2}{2\tau_t^2}=\\
&\log w-\log(1-w)-\frac{\sum^n_{i=1}\phi_{i,t}X_i^2}{2}+\frac{(\eta^\ast_t)^2}{2\sigma_0^2}+\frac{\sum^n_{i=1}\phi_{i,t}(X_i-\eta^\ast_t)^2}{2};
\end{align*}
which holds for any $\eta^\ast_t \in \mathds{R}$. The solutions are given as follows:
\begin{align*}
m_t&=\frac{\sigma_0^2\cdot \sum^n_{i=1}\phi_{i,t}X_i}{\sigma_0^2\cdot \sum^n_{i=1}\phi_{i,t}+1}, t=1,2,\cdots, T\\
\tau_t^2&=\frac{\sigma_0^2}{\sigma_0^2\cdot \sum^n_{i=1}\phi_{i,t}+1}, t=1,2,\cdots,T\\
p_t&=\frac{\exp\left(\log(w_0)-\log(1-w_0)+\log(\sqrt{\sigma_0^2\cdot \sum^n_{i=1}\phi_{i,t}+1})-\frac{\sigma_0^2\cdot (\sum^n_{i=1}\phi_{i,t}X_i)^2}{2(\sigma_0^2\cdot \sum^n_{i=1}\phi_{i,t}+1)}\right)}{\exp\left(\log(w_0)-\log(1-w_0)+\log(\sqrt{\sigma_0^2\cdot \sum^n_{i=1}\phi_{i,t}+1})-\frac{\sigma_0^2\cdot (\sum^n_{i=1}\phi_{i,t}X_i)^2}{2(\sigma_0^2\cdot \sum^n_{i=1}\phi_{i,t}+1)}\right)+1},\\
t&=1,2,\cdots,T.
\end{align*}

Next we deal with the optimal form for $q(\mathbf{V})$, which satisfies
\begin{align*}
\log q(\mathbf{V})&=\mathds{E}_\mathbf{Z}[\log(\prod^{T-1}_{t=1}(1-V_t)^{\alpha_0-1}\cdot V_1^{\sum^n_{i=1}1_{Z_i=1}}\cdot (V_2(1-V_1))^{\sum^n_{i=1}1_{Z_i=2}}\cdots\\ &(V_{T-1}\prod_{t=1}^{T-2}(1-V_t))^{\sum^n_{i=1}1_{Z_i=T-1}}(\prod^{T-1}_{t=1}(1-V_{t}))^{\sum^n_{i=1}1_{Z_i=T}})]+\text{const}\\
&=\sum^n_{i=1}\phi_{i,1}\cdot \log V_1+(\alpha_0-1+\sum^T_{t=2}\sum^n_{i=1}\phi_{i,t})\log (1-V_1)+\sum^n_{i=1}\phi_{i,2}\log V_2+(\alpha_0-1+\\
&\sum^T_{t=3}\sum^n_{i=1}\phi_{i,t})\cdot \log(1-V_2)+\cdots+\sum^{n}_{i=1}\phi_{i,T-1}\cdot \log V_{T-1}\\
&+(\alpha_0-1+\sum^n_{i=1}\phi_{i,t})\log(1-V_{T-1})+\text{const}\\
&\equiv \sum^{T-1}_{t=1}\log q(V_t);
\end{align*}

where $\log q(V_1)=\sum^n_{i=1}\phi_{i,1}\cdot \log V_1+(\alpha_0-1+\sum^T_{t=2}\sum^n_{i=1}\phi_{i,t})\log (1-V_1)+\text{const}$, $\log q(V_2)=\sum^n_{i=1}\phi_{i,2}\log V_2+(\alpha_0-1+\sum^T_{t=3}\sum^n_{i=1}\phi_{i,t})\cdot \log(1-V_2)+\text{const}$, $\cdots$, $\log q(V_{T-1})=\sum^{n}_{i=1}\phi_{i,T-1}\cdot \log V_{T-1}+(\alpha_0-1+\sum^n_{i=1}\phi_{n,T})\log(1-V_{T-1})+\text{const}$. Thus we proved
\[q_{\boldsymbol{\gamma}_1,\boldsymbol{\gamma}_2}(\mathbf{V})=\prod^{T-1}_{t=1}q_{\gamma_{1t},\gamma_{2t}}(V_t).\]
Besides, $V_1$ follows Beta Distribution with parameters $(\gamma_{11},\gamma_{21})=(\sum^n_{i=1}\phi_{k,1}+1,\alpha_0+\sum^T_{t=2}\sum^n_{i=1}\phi_{i,t})$, $V_2$ follows Beta Distribution with parameters $(\gamma_{12},\gamma_{22})=(\sum^n_{i=1}\phi_{k,2}+1,\alpha_0+\sum^T_{t=3}\sum^n_{i=1}\phi_{i,t})$,$\cdots$, $V_{T-1}$ follows Beta Distribution with parameters $(\gamma_{T-1,2},\gamma_{T-1,2})=(\sum^n_{i=1}\phi_{k,T-1}+1,\alpha_0+\sum^n_{i=1}\phi_{i,t})$.

Finally, we deal with $q(\mathbf{Z})$. We have the following optimal form
\begin{align*}
\log q(\mathbf{Z})&=\mathds{E}_{\boldsymbol{\xi},\boldsymbol{\eta^\ast},\mathbf{V}}[\log(\prod^T_{t=1}\pi_t^{\sum^n_{i=1}1_{Z_i=t}})\cdot\exp(-\frac{\sum^n_{i=1}\sum^T_{t=1}(X_i-\eta^\ast_t)^21_{Z_i=t}}{2})]+\text{const}\\
&=\sum^n_{i=1}\mathds{E}_{\boldsymbol{\xi},\boldsymbol{\eta^\ast},\mathbf{V}}[\sum^T_{t=1}(\log \pi_t-\frac{(X_i-\eta^\ast_t)^2}{2})1_{Z_i=t}]+\text{const}\\
&=\sum^n_{i=1}\sum^T_{t=1}(\mathds{E}_{\mathbf{V}}(\log \pi_t)-\frac{\mathds{E}_{\xi_t,\eta^\ast_t}(X_i-\eta^\ast_t)^2}{2})1_{Z_i=t}+\text{const}\\
&=\sum^n_{i=1}\log q(Z_i);
\end{align*}
therefore we proved $q_{\boldsymbol{\Phi}}(\mathbf{Z})=\prod^n_{i=1}q_{\boldsymbol{\phi}_i}(Z_i)$.
Since $\mathds{E}_{\xi_t,\eta^\ast_t}(X_i-\eta^\ast_t)^2=X_i^2-2(1-p_t)m_tX_i+(1-p_t)(m_t^2+\tau^2_t)$, we have
\begin{align*}
\log q(Z_i)&=\sum^T_{t=1}(\mathds{E}_{\mathbf{V}}(\log \pi_t)+(1-p_t)m_tX_i-\frac{1}{2}(1-p_t)(m_t^2+\tau_t^2))1_{Z_i=t}+\text{const}\\
&=\sum^T_{t=1}[\mathds{E}_{\gamma_{1,t},\gamma_{2,t}}(\log V_t)+\sum^{t-1}_{i=1}\mathds{E}_{\gamma_{1,i},\gamma_{2,i}}(\log (1-V_i))\\
&+(1-p_t)m_tX_i-\frac{1}{2}(1-p_t)(m_t^2+\tau_t^2)]1_{Z_i=t}+\text{const}.
\end{align*}

Therefore $q(Z_i)$ is the probability mass function of Multinomial Distribution. Once we fix $i$, $\phi_{i,t}\propto \exp(S_t)$, where $S_t=\exp[\mathds{E}_{\gamma_{1,t},\gamma_{2,t}}(\log V_t)+\sum^{t-1}_{i=1}\mathds{E}_{\gamma_{1,i},\gamma_{2,i}}(\log( 1-V_i))+(1-p_t)m_tX_i-\frac{1}{2}(1-p_t)(m_t^2+\tau_t^2)]$.

\bibliographystyle{chicago}
\bibliography{Needle}
\end{document}